# Raman and Terahertz Spectroscopy of Low-Frequency Chiral Phonons in Amino Acids

Rahul Rao,*,[†] Won Jin Choi,*,[‡] Joseph M. Slocik,[†,¶] Thuc T. Mai,[†,¶] Michael A. Susner,[†] Kelsey A. Collins,[†,^] Michael J. Newburger,[†] Petr Bouř,[§] and Nicholas A. Kotov*,[∥,⊥,#,@]

[†]Materials and Manufacturing Directorate, Air Force Research Laboratory, Wright-Patterson AFB, Ohio 45433, United States

[‡]Physical and Life Sciences, Lawrence Livermore National Laboratory, Livermore, California 94550, United States

[¶]BlueHalo, an AV company, Dayton, Ohio 45433, United States

[^]Core4ce, Dayton, OH 45422, United States

[§]Institute of Organic Chemistry and Biochemistry, Flemingovo nám. 2, 16610 Prague, Czech Republic

[∥]Department of Chemical Engineering, University of Michigan, Ann Arbor, Michigan 48109, United States

[⊥]Biointerfaces Institute, University of Michigan, Ann Arbor, Michigan 48109, United States

[#]Department of Materials Science and Engineering, University of Michigan, Ann Arbor, Michigan 48109, United States

[@]Center for Complex Particle Systems, University of Michigan, Ann Arbor, Michigan 48109, United States

E-mail: rahul.rao.2@us.af.mil; choi21@llnl.gov; kotov@umich.edu

**Abstract:**

Chiral phonons are mirror-symmetric vibrations that correspond to twisting and rotational motions of atoms. In chiral biomolecules, they correspond to low-energy terahertz (THz)-range vibrations of the molecular segments involving dozens of atoms whose energies are sensitive to the chirality of the molecules and local atomic geometries. Here we present spectral signatures of chiral phonons in circularly polarized low-frequency Raman and Raman optical activity (ROA) spectra from crystals of several amino acids in different enantiomeric forms. Along with complementary THz circular dichroism (TCD) measurements, our ROA data reveal two sets of bisignate peaks in valine, alanine, tyrosine and proline between 30 – 150 cm$^{-1}$ (~1 – 4.5 THz) that are more intense than the ROA peaks in the fingerprint region. Density functional theory (DFT) calculations on *L*-alanine attribute these modes to twisting and shearing molecular motions. The strong agreement between the ROA and TCD data demonstrates the power of these complementary vibrational spectroscopy techniques to identify chiral phonons in biomolecules, and offers new insights into their vibrational properties and interactions with circularly polarized light.

**Introduction:**

Chiral objects, i.e. those that cannot be geometrically superimposed on their mirror image,[1] give rise to structural asymmetry between right- and left-handed enantiomers of molecules, nanoparticles, their assemblies, and other chemical structures. In case of crystals, chirality is also associated with the collective vibrational motions of the atoms propagating in them that are known as phonons. While phonons have traditionally been considered to possess only linear momentum, recent experimental and theoretical observations in both magnetic and non-magnetic materials have revealed the existence of chiral phonons, which carry non-zero angular momenta.[2–6] Chiral phonons can be defined as quantized vibrational modes in which atoms or groups of atoms in a solid undergo rotational motion perpendicular to the direction of propagation of the vibration within the lattice. They have been observed using various spectroscopic techniques in enantiomers both of organic and inorganic materials.[6–10]

There are two primary ways to register chiral phonons. Firstly, slight differences in the angular momenta of the incident and scattered phonons result in a small but finite (typically a few cm$^{-1}$) splitting in peak frequencies when excited by right- or left-handed circularly polarized (RCP and LCP, respectively) light.[11] The second approach is to measure optical activity, which is the ability of chiral phonons to rotate polarized light. Two most typical techniques for such measurements are vibrational circular dichroism (VCD) and Raman optical activity (ROA), which are the chiroptical counterparts of infrared absorption spectroscopy and Raman scattering, respectively. VCD relies on the differential absorption of left- and right-circularly polarized light ($I_R - I_L$) during a vibrational transition, while ROA measures the differential intensity of Raman scattered light for RCP and LCP excitation or detection.[12–15] Lately, terahertz (THz) circular dichroism (TCD) and terahertz optical rotation dispersion (TORD) were added to this line-up as the most promising new methodologies to identify chiral phonons in crystals and large molecules.

Typical ROA measurements on organic biomaterials have been limited to the high frequency chemical fingerprint region (800 - 3000 cm$^{-1}$), where the modes correspond to bending and stretching of organic moieties. In contrast, chiral phonons generally correspond to long-range, low-energy vibrations of the crystal lattices assembled from the chiral molecules. These modes occur in the low-frequency regions in the THz part of the spectrum, necessitating the transition from VCD measurements operating in mid- infrared part of the spectrum to TCD.[7,16,17] Recent studies of chiral phonons in multiple chemical structures, including nanostructured microparticles, crystalline nanowires, and microscale crystals, have revealed low-frequency chiral phonons between 1 – 2 THz,[7,17] and although their ROA response might be anticipated, it has not yet been investigated through direct comparison with TCD. While the differences between direct absorption by a THz beam and Raman scattering excited by visible laser light may be related, they may also differ due to distinct selection rules; thus, understanding these variations would be highly significant. Here, we take advantage of the overlap in energies/frequencies between TCD measurements and low-frequency Raman spectroscopy (0 – 4.5 THz, or 0 – 150 cm$^{-1}$), and present

TCD and ROA spectra from low-frequency chiral phonons in enantiomers of amino acid (AA) crystals such as valine, alanine, proline and tyrosine.

We begin with a comparison of THz absorption (TA)/TCD and Raman/ROA spectra from enantiomers of two exemplar AAs, valine and alanine (chemical structures of *D*- and *L*-Val, and *D*- and *L*-Ala are shown in **Figures 1a** and **1b**, respectively), which crystallize in the monoclinic ($P2_1$ space group) and orthorhombic ($P2_12_12_1$ space group) crystal structures, respectively (**Tables S1** and **S2** detail the our single crystal X-ray diffraction refinements for *L*-Val and *L*-Ala, respectively). The TA and TCD measurements were conducted with the help of a custom home-built THz time domain polarimetry setup described in previous studies in Refs.6 and 17. The ROA measurements were conducted with RCP or LCP laser excitation (785 nm), achieved by manually inserting a half waveplate and a quarter waveplate prior to the objective lens in our micro-Raman setup. The quarter waveplates are fixed in place, rotated at 45° to transmit RCP or LCP light. The backscattered light is directed through the same waveplates and through a polarizing beamsplitter before entering the spectrometer (a schematic of our optical layout is shown in **Figure S1**). The scattered light is thus polarized parallel or perpendicular, giving us co- and cross-circularly polarized configurations (RR/LL and RL/LR, respectively). We note that traditional ROA measurements are conducted by continuously modulating the circular polarization of the incident or scattered light (or both). Our spectroscopic set up does not use modulators and is based on continuous accumulation of data for one and the other handedness of the incident light. The set up was validated by collecting cross-circularly polarized ROA spectra from (+)- and (-)-α-pinene (**Figure S2**).

The TA spectra were collected from concentrated mineral oil slurries containing 50 wt.% recrystallized AAs. For both *D*- and *L*-enantiomers of valine and alanine, the TA spectra (0.2 - 5.7 THz) display several sharp absorption peaks between 1 – 4 THz, as shown in **Figures 1a** and **1b**. with frequency ranges indicated the top axes. These results are consistent with previous far-IR spectra reported for *L*-Val and *L*-Ala.[19] Circularly polarized Raman spectra in the low-frequency region were obtained from the AA crystals deposited directly from solution onto silicon substrates and measured under cross-circularly polarized conditions with RL and LR configurations, as displayed in **Figure 1**. Similar to the TA spectra, the Raman spectra exhibit multiple peaks between 10 – 200 cm$^{-1}$, with peaks from both enantiomers appearing at the same frequencies. These findings are in agreement with previously published low-frequency Raman data on AAs.[20–22] Notably, the correspondence between the TA and Raman spectra from valine is strong: below 100 cm$^{-1}$ (3 THz), the two most intense Raman modes at 55 and 100 cm$^{-1}$ align well with the two most intense peaks in the TA spectra at 1.68 and 2.8 THz, respectively.

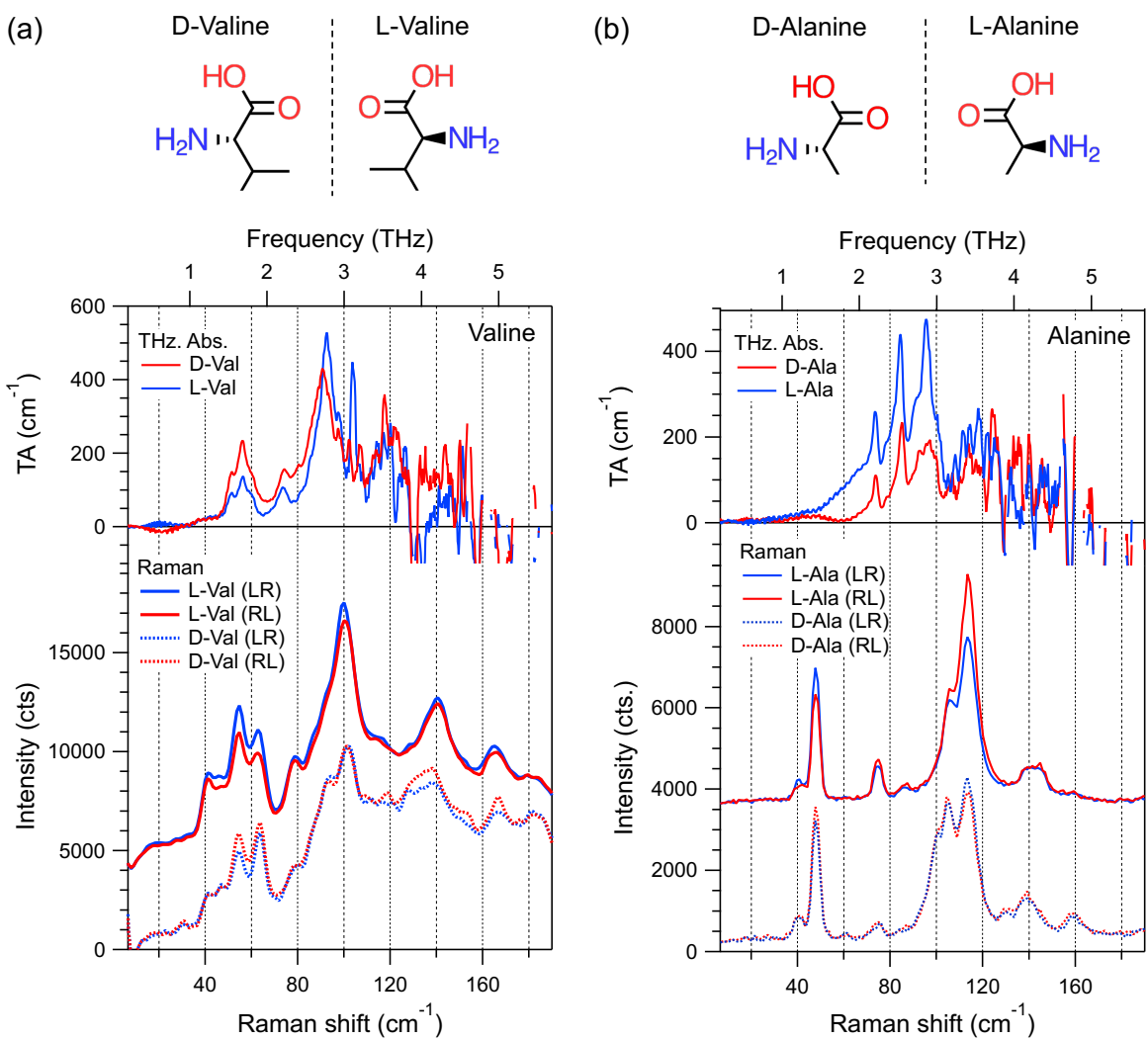

**Figure 1**. Circularly polarized TA and low-frequency Raman spectra with frequencies corresponding to the top and bottom axes, respectively: (a) *D*- and *L*-Val, and (b) *D*- and *L*-Ala crystals. The molecular structures of the two AA enantiomers are shown in the schematics on top.

We also observed variations between the TA and Raman spectra in the number of peaks and their frequencies, which can be attributed to the intrinsic difference in the selection rules between the two techniques. As in case of conventional infrared spectroscopy, the molecular vibrations in TA measurements must cause a net change in their dipole moment. On the contrary, Raman-active vibrations result from a change in the polarizability of a molecule**.** The differences between the spectra obtained from the two techniques can be seen more starkly in *D*- and *L*-Ala (**Figure 1b**), where the most intense peaks in the Raman spectra appear around 48 and 113 cm$^{-1}$ (1.4 and 3.4 THz), whereas the most intense features in the TA spectra lie between the two frequencies (2.5 – 3 THz). These peaks between 2.5 and 3 THz do appear as weak features in the Raman spectra. Their appearance can be potentially attributed to differences between the TA and Raman measurements (spot sizes and penetration depths) as well as selection rules. `Note that the Raman spectra can also access frequencies above 3 THz (100 cm$^{-1}$), allowing for the observation of more extended range structural vibrational modes in the low-frequency range.

The origin of the low-frequency peaks in the TA and Raman spectra can be attributed to long-range rotary motions of the chemical groups in the structure of the AA enantiomers. The $P2_1$ and $P2_12_12_1$ space groups contain 1 and 3 screw axes, respectively, that produce chiral patterns within the AA unit cells. This results in nitrogen atoms of the amino groups (–$NH_2$) and N–H···O hydrogen bonds forming helices within each cell. To show the effects of the structural chirality on the vibrational modes, we present TCD and ROA spectra from the two AAs in **Figure 2**. *D*- and *L*-Val have classical bisignate shapes between 1-2 THz in their TCD spectra (top panel, **Figure 2a**). The corresponding features in alanine are less clear. However, the ROA spectra ($I_{RL} - I_{LR}$, written as $I_R - I_L$ for brevity) for both AAs exhibit strong bisignate peaks. Note that ROA intensities in our measurements are higher than the typically measured ROA spectra. We attribute the higher intensities to the differences in our measurement setup versus the traditional ROA techniques. Nonetheless, the opposite intensities of these peaks in the ROA spectra are consistent with their assignment as chiral phonons. We see these peaks in two frequency regions, between 40 – 60 cm$^{-1}$ and the second one between 100 – 140 cm$^{-1}$ (bottom panels in **Figures 2a** and **2b** for valine and alanine, respectively). The bisignate ROA spectra can also be seen in the fingerprint regions of the spectra from both valine and alanine (**Figure S3**), and agree well with previous reports.[14,23]

Other than valine and alanine, we also observe bisignate ROA peaks in the low-frequency range from enantiomers of tyrosine and proline (**Figure S4**). Tyrosine exhibits a strong and clear chiral phonon signature at the same frequency (32 cm$^{-1}$, or 0.96 THz) in both TCD and ROA spectra, but the peaks in proline are broader, with less of a correspondence between the TCD and ROA spectra. Interestingly, for a given enantiomer, some of the ROA peaks exhibit opposite intensities between the AAs. These contrasts could be attributed to differences in crystallization and/or crystallite orientations within the polycrystalline samples measured in this study.

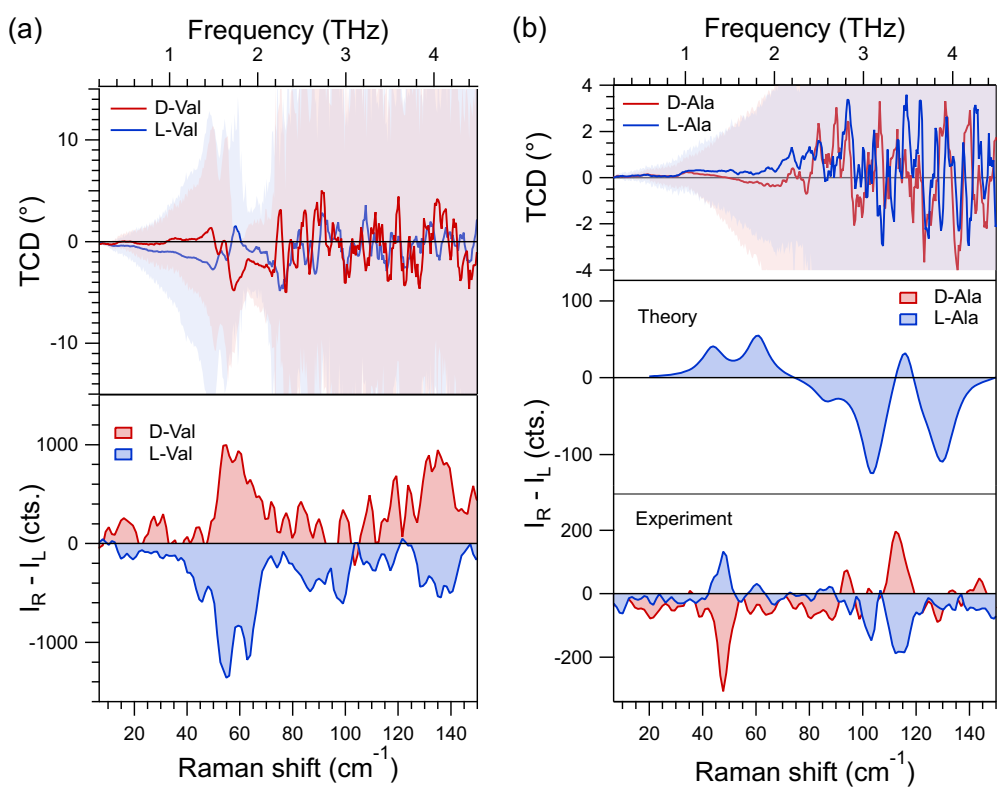

**Figure 2**. TCD and ROA spectra with frequencies corresponding to the top and bottom axes, respectively from *D* and *L*-enantiomers of (a) Valine and (b) Alanine. The ROA spectrum for *L*-Ala calculated by DFT is also plotted in the middle panel of (b).

To gain further insights into the origin of the low-frequency Raman modes, we calculated the Raman and ROA spectrum of *L*-Ala using DFT (see Methods). The theoretical Raman spectrum matches well with the measured spectrum across both the low-frequency and fingerprint regions (**Figure S4**). The Raman mode calculations were performed by considering an isotropic (polycrystalline) crystal as well as along the *a*, *b* and *c* crystal axes. The calculation reveals four Raman modes in the low-frequency region, at 43, 60, 103 and 130 cm$^{-1}$, which are slightly shifted compared to our experimentally observed peaks at 44, 52, 80, 110 and 118 cm$^{-1}$. Minor discrepancies in frequencies can be attributed to the DFT error and anharmonic effects. Importantly, the calculated ROA spectrum for *L*-Ala (middle panel in **Figure 2b**) matches the experimental spectrum very well. It exhibits positive and negative peaks in the 40-60 cm$^{-1}$ and 100-140 cm$^{-1}$ ranges, respectively, similar to the measured circularly polarized spectrum (bottom panel in **Figure 2b**). While we show the calculated ROA for a polycrystal of *L*-Ala in **Figure 2b**, the spectra along each of the crystallographic axes exhibit ROA peaks with similar signs but varying intensities. All spectra correlate well with the measured ROA spectrum (**Figure S5**).

The calculated Raman modes at 43, 60, 103 and 130 cm$^{-1}$ correspond to twisting and rotary motions that are in opposing directions for the two enantiomers and are characteristic of chiral phonons. The eigenvectors for two of the modes at 60 and 130 cm$^{-1}$ are shown in **Figure 3**. The eigenvectors for the other two modes, namely at 43 and 103 cm$^{-1}$, are shown in the Supporting Information (**Figures S6a** and **S6b**), including animations of all four mode vibrations. Note that the schematics in **Figures 3** and **S6** show the zwitterionic form of L-Ala. The modes at 43 and 60 cm$^{-1}$ primarily involve shearing motions of the molecules within the alanine unit cell, accompanied by a small amount of twisting motion. This can be seen more clearly in **Figure S6c**, which shows a rotated view of the *L*-Ala unit cell. The 103 and 130 cm$^{-1}$ modes primarily involve twisting of the carboxylate and methyl groups in opposite directions. We also calculated the TA spectrum for *L*-Ala (**Figure S7**), which shows a good match with the experimentally observed peaks at 1.4 and 1.8 THz. The calculations are consistent with previous work that showed that the 1.2–1.4 THz peak in the TA spectrum from *L*-glutamine involves the twisting of the carboxylate groups in both the main and side chains.[7]

We note that, while we observed differences in intensities in the chiral phonon modes with RCP or LCP excitation, we do not see any splitting in peak frequencies. One reason could be the presence of several overlapping vibrational modes with closely spaced frequencies as revealed by the calculations (**Figure S5**). Another reason could be that the splitting is present and finite, but is small enough that it is below our instrument resolution of 1 cm$^{-1}$. Indeed, such a small spitting and opposite angular momenta were recently calculated for an achiral two-dimensional material AgCrP$_2$Se$_6$,[24] which did not exhibit any splitting of frequencies in room temperature Raman spectra, but nonetheless exhibited ROA for several low-energy Raman modes. ROA spectra with negligible or very small frequency splitting have also been reported for other achiral materials such as TaS$_2$, ReS$_2$ and ReSe$_2$ and chiral materials like AgClO$_3$ and hybrid organic inorganic perovskites.[25–31]

In summary, we have presented TA, TCD, Raman and ROA spectra from AA crystals, showing several peaks in the low-frequency region (below 150 cm$^{-1}$ or 4.5 THz). The ROA data show clear bisignate peaks arising from chiral vibrational modes, with intensities greater than those in the fingerprint region. Computations of the Raman and ROA spectrum of *L*-alanine reveal several low-frequency Raman-active modes corresponding to shearing and twisting of the alanine molecules within the unit cell, as well as twisting of the carboxylate and methyl groups. Overall, our study highlights the strong influence of structural chirality on the interaction between chiral phonons and circularly polarized light, and demonstrates low-frequency polarized Raman spectroscopy as a valuable tool for identifying low-energy chiral phonons in organic materials.

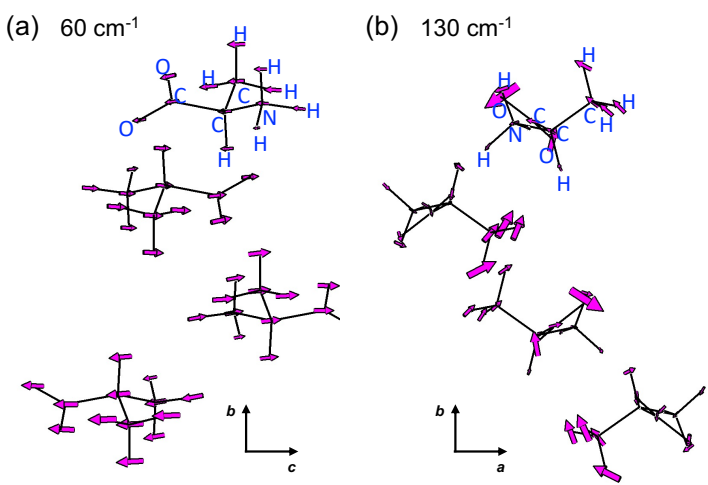

**Figure 3**. Schematics of the eigenvectors for the two calculated Raman modes at (a) 60 cm$^{-1}$ and (b) 130 cm$^{-1}$ in the L-Ala unit cell. The atoms in the alanine molecule are labeled for the top molecule in the unit cell.

**Methods**

**TA and TCD measurements:** Terahertz absorption (TA) and terahertz circular dichroism (TCD) spectroscopies are complementary chiroptical methods used to probe collective vibrational modes—particularly chiral phonons—in molecular and biomolecular crystals. TA spectroscopy measures the frequency-dependent absorption of terahertz radiation as it passes through a sample, revealing resonance peaks corresponding to phononic or intermolecular vibrations within the 0.2–3 THz range (6–100 cm$^{-1}$). Briefly, for the TA and TCD measurements, we used THz time-domain polarimetry based on a 1550 nm fiber-coupled femtosecond laser (>500 mW, 100 MHz) with a TERA15-TX-FC antenna (Menlo Systems) as the THz emitter. For detection, a delayed probe beam was used with a TERA15-RX-FC antenna (Menlo Systems), and the voltage signal was amplified. The system employed four TPX lenses (f = 50 mm) to collimate the THz beam and focus it at the sample position, producing a focal spot of ~500 µm in diameter at 1 THz, with a depth of focus of roughly 1 mm. Three THz linear polarizers (Microtech, G30-s) were used: the first and third, placed directly in front of the THz emitter and detector, ensured linearly polarized light, while the second polarizer was rotated to measure TCD and TORD. For sample preparation, all amino acid samples were purchased from Sigma-Aldrich and ground into small crystals using a mortar and pestle. The powders were then mixed with mineral oil to form a slurry. The sample slurries were spread onto a quartz slide and sandwiched with another quartz slide. The slurry thickness was controlled using a spacer; in this case, we used 3M double-sided tape.

**Circularly polarized Raman spectroscopy:** Room temperature circularly polarized Raman spectra were collected in a Renishaw inVia Raman microscope. Our instrument is outfitted with a low-frequency module (Coherent/Ondax THz Raman probe) that uses fiber optics to couple a 785 nm laser into the objective lens for excitation and to direct the scattered light into the inVia spectrometer. The optical layout for achieving circular polarization is shown in the **Figure S1**. L- and D-amino acids were purchased from Sigma Aldrich at >98% purity and dissolved in doubly deionized water to yield concentrations of 10 mg/mL. 2 µL of each amino acid enantiomer was spotted onto a single-sided polished silicon wafer and allowed to air-dry at room temperature for 2 hours. All spectra were collected by focusing the excitation laser on the AA crystals through a 50x objective lens. The laser power used for the spectral collection was 1.8 mW, and spectra were collected with a 10 s exposure time and 12 accumulations. At least 3 spectra were collected from different spots and averaged to obtain the LR and RL spectra.

**Single Crystal X-ray diffraction:** X-ray diffraction measurements were performed with a Rigaku KtaLAB synergy-I single crystal diffractometer with Cu k$_\alpha$ radiation ($\lambda$ = 1.5406 Å). Refinements were completed with the onboard Rigaku CrysAlis Pro software using ShelX[32] and are presented in the Supplementary Information table S1 and S2 for R-Val and L-Ala, respectively.

**Raman and ROA calculations:** The crystal cell and geometry of L-alanine crystal were first optimized by energy minimization using the CASTEP software.[33] The X-ray structure was used as the initial geometry (number 278466 in the Cambridge Crystallographic Data Centre), the PBE functional was used with 700 eV basis set cut-off, which provided lattice cell parameters (5,88, 11.86 and 5.73 Å), comparable with the lattice parameters obtained experimentally (5.94, 12.26, 5.79 Å). The harmonic force field (dynamic matrix) and the zero-phonon eigenvectors were obtained at the same level as the optimization. To get polarizability derivatives needed for Raman and ROA,[34] a cluster of alanine molecule and nearest neighbors was partially optimized in the vibrational normal mode coordinates[35] and the Gaussian program[36] was used, applying the B3LYP/6-311++G**/PCM method. The polarizability derivatives were then transferred on the lattice cell geometry using the Cartesian coordinate-based tensor transfer[37] and backscattered Raman and ROA intensities for the scattered circular polarization (SCP) were generated. Smooth spectra were obtained by a convolution of the intensities with Lorentzian functions, using a bandwidth of 3 or 10 cm$^{-1}$.

**Acknowledgements**

R.R., T.T.M and M.A.S acknowledge funding from the Air Force Office of Scientific Research (AFOSR) LRIR grant no. 23RXCOR003. N.A.K. and W.C are grateful for the financial support from National Science Foundation (NSF) grant #2243104, *Center for Complex Particle Systems (COMPASS)*; grant #2317423 *Lock-and-Key Interactions of Proteins and Chiral Nanoparticles*, grant #2418861 CBET-EPSRC *Chiroptical Second-Harmonic Scattering of Nanostructures and Their*

*Biocomplexes*. The authors also acknowledge an essential support from European Research Council via collaborative Synergy grant: 101166855 - GAP-101166855, CHIRAL-PRO: *Handshake Complexes of Chiral Nanoparticles and Proteins*. P.B acknowledges ssupport from the Grant Agency of the Czech Republic (24-10558S).